\documentclass[a4paper,aps,prd,10pt,preprintnumbers,showpacs,twocolumn,superscriptaddress,nofootinbib,amsmath,amssymb]{revtex4-1}
\usepackage{graphicx}
\usepackage{cmap}
\usepackage[utf8]{inputenc}
\usepackage[T1]{fontenc}

\def\re#1{Re(#1)}

\def\K{{\cal K}}
\def\Order#1{{\cal O}\left(#1\right)}

\begin{document}

\title{Infinite tower of higher-curvature corrections: \\ Quasinormal modes and late-time behavior of D-dimensional regular black holes}

\author{R. A. Konoplya}
\email{roman.konoplya@gmail.com}
\affiliation{Research Centre for Theoretical Physics and Astrophysics, Institute of Physics, Silesian University in Opava, Bezručovo náměstí 13, CZ-74601 Opava, Czech Republic}

\author{A. Zhidenko}
\email{olexandr.zhydenko@ufabc.edu.br}
\affiliation{Centro de Matemática, Computação e Cognição (CMCC), Universidade Federal do ABC (UFABC), \\ Rua Abolição, CEP: 09210-180, Santo André, SP, Brazil}

\begin{abstract}
Recently, Bueno, Cano, and Hennigar [arXiv:2403.04827] proposed a generic approach for incorporating an infinite tower of higher-curvature corrections into the Einstein theory. In this study, we compute quasinormal modes for certain regular $D$-dimensional black holes resulting from this infinite series of higher-curvature corrections, specifically focusing on the $D$-dimensional extensions of the Bardeen and Hayward black holes. We demonstrate that while the fundamental mode is minimally affected by moderate coupling constants, the higher overtones exhibit significant sensitivity even to small coupling values, yielding unconventional modes characterized by vanishing real oscillation frequencies. When comparing the frequencies derived from the metric truncated at several orders of higher-curvature corrections with those resulting from the infinite series of terms, we observe a rapid convergence of the frequencies to their limit for the complete regular black hole. This validates the extensive research conducted on specific theories with a finite number of higher-curvature corrections, such as the Lovelock theory.
\end{abstract}

\pacs{04.30.Nk,04.50.Kd,04.70.Bw}
\maketitle

\section{Introduction}

The black-hole solution of the vacuum Einstein equations possesses singularity at the origin.
A number of attempts were made to resolve the singularity problem via adding some, frequently not well-motivated, matter, to the gravitational sector, such as various types of exotic matter (including nonexisting nonlinear electrodynamics), stress-energy tensors violating energy conditions, etc. \cite{Sakharov:1966aja,Bardeen,Dymnikova:1992ux,Borde:1994ai,Hayward:2005gi,Lemos:2011dq,Bambi:2013ufa,Simpson:2018tsi,Rodrigues:2018bdc,Mars:1996khm,Ayon-Beato:1998hmi,Bronnikov:2000vy,Ayon-Beato:2000mjt,Bronnikov:2000yz,Ayon-Beato:2004ywd,Dymnikova:2004zc,Berej:2006cc,Balart:2014jia,Fan:2016rih,Bronnikov:2017sgg,Junior:2023ixh,Ansoldi:2008jw,Lemos:2011dq,Sajadi:2017glu}.

Recently new families of regular black holes in $D \geq 5$ spacetime dimensions have been obtained as a result of inclusion of the infinite number of higher-curvature terms correcting the Einstein action \cite{Bueno:2024dgm}. This approach is part of the quasi-topological gravity \cite{Oliva:2010eb,Myers:2010ru,Dehghani:2011vu,Ahmed:2017jod,Cisterna:2017umf}.
The privilege of this approach is that no exotic or poorly motivated state of matter is introduced to produce a singularity free black hole.
The obtained regular black holes \cite{Bueno:2024dgm} are generic regular, static, and spherically symmetric solutions of the corresponding theories.

Here we consider two of the aforementioned regular black-hole solutions, also because their four-dimensional analogs were derived within different contexts: the Hayward black hole \cite{Hayward:2005gi} and the Bardeen-like black hole \cite{Bardeen}. The four-dimensional Hayward metric was derived within the asymptotically safe gravity \cite{Held:2019xde} and, initially as a model for a quantum corrected evaporating black hole \cite{Hayward:2005gi}. The Bardeen metric was initially suggested as an ad hoc solution, but, apart from exotic electrodynamics, cited above, it can describe a quantum corrected black hole found either via corrections to the black-hole thermodynamics \cite{Maluf:2018ksj} or via T-duality \cite{Nicolini:2019irw}.

One of the questions we aim to address is the extent to which truncation of the infinite series at the initial few orders, as occurs, for example, in the Lovelock theory, is justified. If the fundamental characteristics of black holes in the truncated theory significantly differ from those in theories with an infinite number of higher-curvature corrections, then studying the truncated theory at a lower order yields little, if any, meaningful information.

One of such basic characteristics of black holes is its set of proper oscillation frequencies, called \emph{quasinormal modes}.
They are important not only from the point of view of observation of radiation phenomena around black holes \cite{LIGOScientific:2016aoc}, but also for testing stability of fields under consideration \cite{Konoplya:2011qq}.
The quasinormal modes of the four-dimensional Hayward and Bardeen black holes have been extensively studied in the literature (see~\cite{Konoplya:2023ahd,Konoplya:2023ppx,Fernando:2012yw,Flachi:2012nv,Toshmatov:2019gxg,Nomura:2020tpc,Franzin:2023slm,Breton:2016mqh,Liu:2020ddo,Zhang:2024nny,Bolokhov:2023ruj} and references therein).
Here we attack scalar and electromagnetic perturbations of the $D$-dimensional generalizations of the Hayward and Bardeen black holes. Special attention is devoted to the overtones' behavior, because, as was shown in \cite{Konoplya:2022pbc,Konoplya:2023hqb}, relatively small near-horizon corrections could hardly lead to a strong change of the fundamental mode that is localized near the peak of the effective potential, but, on the contrary, the first few overtones are extremely sensitive to the least near horizon deformations. This phenomenon was called \emph{outburst of overtones} and extensively studied in several recent publications \cite{Gong:2023ghh,Bolokhov:2023bwm,Bolokhov:2023ruj,Konoplya:2023kem,Konoplya:2023aph,Konoplya:2022iyn}.

We will show that the fundamental mode is greatly affected by the coupling parameter for both models, once the coupling constant is increased up to the extreme value supporting the existence of the event horizon. What is more, the first few overtones of the Hayward $D$-dimensional black hole change a lot even for relatively small values of the coupling, and the second and higher overtones (depending on $D$) show peculiar behavior: The real oscillation frequency quickly goes to zero, once the coupling is turned on.

Simultaneously, we demonstrate that, if we take into account a finite number of curvature-correction terms, the fundamental mode rapidly converges to its value in the limit of an infinite number of higher-curvature corrections, thereby affirmatively addressing the earlier question regarding the justification of lower order theories.

Our work is organized as follows. Section~\ref{sec:equations} introduces the basic equations on the metric, perturbation equations and effective potentials. Section~\ref{sec:methods} describes the numerical and semianalytic methods used for calculations of quasinormal modes and asymptotic tails. Sections~\ref{sec:Hayward}~and~\ref{sec:Bardeen} summarize the obtained numerical results on quasinormal modes of the $D$-dimensional regular black holes, and asymptotic tails are considered in Section~\ref{sec:tails}. In Section~\ref{sec:conclusions}, we discuss the obtained results and mention possible new directions of study.

\section{The basic equations}\label{sec:equations}

Following Bueno, Cano, and Hennigar~\cite{Bueno:2024dgm}, the action of the Einsteinian theory with general higher-curvature corrections has the form
\begin{equation}\label{QTaction}
I_{\rm QT}=\frac{1}{16\pi G} \int \mathrm{d}^Dx \sqrt{|g|} \left[R+\sum_{n=2}^{n_{\rm max}} \alpha_n \mathcal{Z}_n \right]\, ,
\end{equation}
where $\alpha_n$ are arbitrary coupling constants with dimensions of length$^{2(n-1)}$ and $\mathcal{Z}_n$ are the Quasi-topological densities \cite{Bueno:2019ycr}.

The spherically symmetric $D$-dimensional black hole is given by the following metric:
\begin{eqnarray}\label{metric}
ds^2 &=& - f(r) dt^2 + \frac{dr^2}{f(r)} + r^2 d\Omega^2_{D-2} \, ,
\\\nonumber &&f(r)\equiv1-r^2\psi(r)\, ,
\end{eqnarray}
such that $\psi(r)$ satisfies
\begin{equation}\label{hequation}
h(\psi(r)) = \frac{\mu}{r^{D-1}}\, ,
\end{equation}
where $\mu$ is the positive constant proportional to the ADM mass.

The function $h(\psi)$ is given by the series,
\begin{equation}\label{hseries}
h(\psi) \equiv \psi + \sum_{n=2}^{n_{\rm max}} \alpha_n \psi^n\, .
\end{equation}

For an infinite tower of higher-curvature corrections ($n_{\rm max}\to\infty$), Bueno, Cano, and Hennigar~\cite{Bueno:2024dgm} have proposed the smooth monotonic functions, which allow for analytic expressions for the metric functions.

The general relativistic equations for a scalar ($\Phi$) and electromagnetic ($A_\mu$) fields can be written as follows:
\begin{subequations}\label{coveqs}
\begin{eqnarray}\label{KGg}
\frac{1}{\sqrt{-g}}\partial_\mu \left(\sqrt{-g}g^{\mu \nu}\partial_\nu\Phi\right)&=&0,
\\\label{EmagEq}
\frac{1}{\sqrt{-g}}\partial_{\mu} \left(F_{\rho\sigma}g^{\rho \nu}g^{\sigma \mu}\sqrt{-g}\right)&=&0,
\end{eqnarray}
\end{subequations}
where $F_{\mu\nu}=\partial_\mu A_\nu-\partial_\nu A_\mu$ is the electromagnetic tensor.

After separation of variables, equations for the test massless fields in the background of the spherically symmetric black hole can be reduced to the wavelike form
\begin{equation}\label{wavelike}
\frac{d^2\Psi}{dr_*^2}+(\omega^2-V(r_*))\Psi(r_*)=0,
\end{equation}
where the tortoise coordinate is defined as follows:
\begin{equation}\label{tortoise}
dr_*=\frac{dr}{f(r)}.
\end{equation}

The effective potentials for the scalar ($V_0$) and electromagnetic ($V_1$ and $V_2$) \cite{Crispino:2000jx,Lopez-Ortega:2006vjp} fields
\begin{subequations}\label{potentials}
\begin{eqnarray}
V_0(r) &=& f(r)\Biggl(\frac{\ell(\ell+D-3)}{r^2} \\\nonumber&&+ \frac{(D-2)(D-4)}{4r^2}f(r) + \frac{D-2}{2r}\frac{df}{dr}\Biggr);
\\
V_1(r) &=& f(r)\Biggl(\frac{\ell(\ell+D-3)}{r^2} \\\nonumber&&+ \frac{(D-2)(D-4)}{4r^2}f(r) - \frac{D-4}{2r}\frac{df}{dr}\Biggr);
\\
V_2(r) &=& f(r)\Biggl(\frac{(\ell+1)(\ell+D-4)}{r^2} \\\nonumber&&+ \frac{(D-4)(D-6)}{4r^2}f(r) + \frac{D-4}{2r}\frac{df}{dr}\Biggr).
\end{eqnarray}
\end{subequations}

\section{Methods used for finding of quasinormal modes and asymptotic tails}\label{sec:methods}

Quasinormal modes of asymptotically flat black holes are frequencies for which the corresponding wave functions satisfy the particular boundary conditions: purely ingoing waves at the event horizon and purely outgoing ones at infinity. The methods for finding quasinormal modes have been reviewed in a great number of papers. Therefore, here, we will only briefly sketch the main properties of them.

\subsection{WKB method}

The semianalytic WKB method is based on the expansion of the wave function in the WKB series at some order at the two asymptotic regions: infinity and the event horizon and matching of these WKB asymptotic solutions with the Taylor expansion near the peak of the effective potential. Thus, we imply the two turning points and a single maximum of the effective potential for an effective application of the WKB approach. The general WKB formula \cite{Schutz:1985km,Iyer:1986np,Konoplya:2003ii,Matyjasek:2017psv}
\begin{eqnarray}\label{WKBformula-spherical}
\omega^2&=&V_0+A_2(\K^2)+A_4(\K^2)+A_6(\K^2)+\ldots\\\nonumber&-&i\K\sqrt{-2V_2}\left(1+A_3(\K^2)+A_5(\K^2)+A_7(\K^2)\ldots\right),
\end{eqnarray}
where for quasinormal modes we have
\begin{equation}
\K=n+\frac{1}{2}, \quad n=0,1,2,\ldots,
\end{equation}
was further improved by using the Padé approximants \cite{Matyjasek:2017psv}. Here we will use the sixth WKB order and the Padé splitting with $\tilde{m}=4$ because it shows the best accuracy for the Schwarzschild limit. We use the Mathematica WKB code shared in \cite{Konoplya:2019hlu}.
We will see that the WKB method with Padé approximants is sufficiently accurate for finding the fundamental mode even at $\ell=0$ case, while the usual WKB formula is usually reasonably accurate at the sixth order only for $\ell > n$. This observation is in concordance with numerous applications of the WKB method and comparison of the results with those obtained by time-domain integration and Frobenius techniques (see, for instance, \cite{Aragon:2020tvq,Bolokhov:2023ruj,Konoplya:2020bxa} and references therein).

\subsection{Time-domain integration}

In order to study asymptotic tails, we will use the time-domain integration method suggested by Gundlach and co-workers \cite{Gundlach:1993tp}. The appropriate discretization scheme is
\begin{eqnarray}
\Psi\left(N\right)&=&\Psi\left(W\right)+\Psi\left(E\right)-\Psi\left(S\right)\label{Discretization}\\
&&- \Delta^2V\left(S\right)\frac{\Psi\left(W\right)+\Psi\left(E\right)}{4}+{\cal O}\left(\Delta^4\right),\nonumber
\end{eqnarray}
The points are designated as follows: $N\equiv\left(u+\Delta,v+\Delta\right)$, $W\equiv\left(u+\Delta,v\right)$, $E\equiv\left(u,v+\Delta\right)$, and $S\equiv\left(u,v\right)$. Initial Gaussian data are imposed on the two null surfaces $u=u_0$ and $v=v_0$. When using the Prony method, this approach allows one to find also the fundamental mode for $\ell>0$ with reasonable accuracy. It is essential that in the time-domain integration method all overtones are taken into consideration, which makes it possible to judge about the stability of the perturbation.

\subsection{Frobenius method}

We will use the Frobenius method, suggested by E. Leaver~\cite{Leaver:1985ax} for finding quasinormal modes, which is based on the convergent series expansion and providing, thereby, precise calculation of quasinormal modes. For quicker convergence, we use the Nollert improvement of the Leaver procedure \citep{Nollert:1993zz,Zhidenko:2006rs}.

The wavelike equation has a regular singular point at the event horizon $r=r_0$ and the irregular singular point at infinity. We introduce the new function $y(r)$ in such a way
\begin{equation}\label{reg}
\Psi(r)= P (r, \omega) y(r),
\end{equation}
that the factor $P(r, \omega)$ provides regularity of $y(r)$ in the whole range from the event horizon to infinity, once the quasinormal modes boundary conditions are fulfilled.
Then, $y(r)$ can be written as the following series:
\begin{equation}\label{Frobenius}
y(r)=\sum_{k=0}^{\infty}a_k\left(1-\frac{r_0}{r}\right)^k.
\end{equation}
Further, we find the recurrence relation for the coefficients $a_k$ and, after employing the Gaussian eliminations, obtain the nonalgebraic equation with infinite continued fraction, which is solved numerically.
It is essential that, when the convergence radius for the series (\ref{Frobenius}) is insufficient, we employ analytic continuation of the function though a sequence of positive real midpoints using the approach developed in \cite{Rostworowski:2006bp}. For an initial guess for the quasinormal frequency, we use the value found by the WKB or time-domain integration approaches.

\section{Quasinormal modes of the D-dimensional Hayward black holes}\label{sec:Hayward}

The $D$-dimensional generalization of the Hayward black hole is the solution for the following choice of the function $h(\psi)$ \cite{Bueno:2024dgm}:
\begin{equation}
h(\psi)=\dfrac{\psi}{1-\alpha \psi},
\end{equation}
which corresponds to the following values of the coefficients in (\ref{hseries}):
$$\alpha_n=\alpha^{n-1}.$$

Then, Eq.~(\ref{hequation}) can be solved with respect to the metric function,
\begin{equation}\label{Haywardmetric}
f(r)=1-\dfrac{\mu r^2}{r^{D-1}+\alpha\mu}.
\end{equation}

In order to measure all quantities in units of the event horizon, we fix the constant $\mu$ as follows:
\begin{equation}
\mu=\dfrac{r_0^{D-1}}{r_0^2 - \alpha}, \qquad 0\leq\alpha\leq\dfrac{D-3}{D-1}r_0^2.
\end{equation}

\begin{table*}
\begin{tabular*}{\textwidth}{@{\extracolsep{\fill}} l c c c c}
\hline
$\alpha$ & scalar field ($\ell=0$) & scalar field ($\ell=1$) & $V_1$ ($\ell=1$) & $V_2$ ($\ell=1$) \\
\hline
$0  $  & $0.533313-0.383482 i$ & $1.016023-0.362324 i$ & $0.753155-0.317570 i$ & $0.952729-0.350732 i$ \\
$0.1$  & $0.513522-0.344841 i$ & $0.975694-0.328308 i$ & $0.736623-0.286883 i$ & $0.922189-0.317769 i$ \\
$0.2$  & $0.485086-0.307249 i$ & $0.927288-0.295652 i$ & $0.709222-0.256504 i$ & $0.882942-0.284722 i$ \\
$0.3$  & $0.447699-0.276072 i$ & $0.870875-0.266048 i$ & $0.672309-0.227728 i$ & $0.833939-0.253698 i$ \\
$0.4$  & $0.410762-0.252623 i$ & $0.807429-0.240295 i$ & $0.625649-0.202608 i$ & $0.775548-0.226912 i$ \\
$0.49$ & $0.377823-0.232288 i$ & $0.744662-0.219755 i$ & $0.577222-0.183931 i$ & $0.715849-0.206776 i$ \\
\hline
\end{tabular*}
\caption{Dominant ($n=0$) quasinormal modes of the scalar and electromagnetic fields for the Hayward black hole ($D=5$, $r_0=1$) calculated using the sixth order WKB formulas with the Padé approximant ($\widetilde{m}=4$).}\label{table:Haywardb}
\end{table*}

\begin{table*}
\begin{tabular*}{\textwidth}{@{\extracolsep{\fill}} l c c c c}
\hline
$\alpha$ & scalar field ($\ell=0$) & scalar field ($\ell=1$) & $V_1$ ($\ell=1$) & $V_2$ ($\ell=1$) \\
\hline
$0   $ & $0.889368-0.532938 i$ & $1.446598-0.509220 i$ & $1.047467-0.434546 i$ & $1.400026-0.498177 i$ \\
$0.1 $ & $0.865535-0.492948 i$ & $1.406060-0.473736 i$ & $1.017864-0.408643 i$ & $1.367990-0.463403 i$ \\
$0.2 $ & $0.832881-0.453052 i$ & $1.357627-0.439019 i$ & $0.990503-0.380849 i$ & $1.327738-0.428081 i$ \\
$0.3 $ & $0.791088-0.417394 i$ & $1.301068-0.406335 i$ & $0.955109-0.350414 i$ & $1.278185-0.393560 i$ \\
$0.4 $ & $0.745566-0.389018 i$ & $1.236510-0.376336 i$ & $0.911101-0.320106 i$ & $1.218611-0.361360 i$ \\
$0.5 $ & $0.698387-0.363259 i$ & $1.163511-0.348625 i$ & $0.858389-0.292015 i$ & $1.148538-0.332429 i$ \\
$0.59$ & $0.652794-0.339172 i$ & $1.088942-0.324592 i$ & $0.803432-0.269975 i$ & $1.075395-0.308741 i$ \\
\hline
\end{tabular*}
\caption{Dominant ($n=0$) quasinormal modes of the scalar and electromagnetic fields for the Hayward black hole ($D=6$, $r_0=1$) calculated using the sixth order WKB formulas with the Padé approximant ($\widetilde{m}=4$).}
\end{table*}

\begin{table*}
\begin{tabular*}{\textwidth}{@{\extracolsep{\fill}} l c c c c}
\hline
$\alpha$ & scalar field ($\ell=0$) & scalar field ($\ell=1$) & $V_1$ ($\ell=1$) & $V_2$ ($\ell=1$) \\
\hline
$0   $ & $1.270695-0.665502 i$ & $1.881482-0.641045 i$ & $1.396098-0.525823 i$ & $1.845853-0.630962 i$ \\
$0.1 $ & $1.243472-0.624555 i$ & $1.839963-0.604535 i$ & $1.367197-0.495529 i$ & $1.812011-0.594948 i$ \\
$0.2 $ & $1.207611-0.583608 i$ & $1.790642-0.568601 i$ & $1.315762-0.484553 i$ & $1.770205-0.558289 i$ \\
$0.3 $ & $1.162657-0.546125 i$ & $1.733106-0.534344 i$ & $1.284492-0.461881 i$ & $1.719313-0.522050 i$ \\
$0.4 $ & $1.111858-0.514869 i$ & $1.667177-0.502260 i$ & $1.241617-0.431980 i$ & $1.658335-0.487314 i$ \\
$0.5 $ & $1.057286-0.486705 i$ & $1.591965-0.471912 i$ & $1.187389-0.401239 i$ & $1.586099-0.454785 i$ \\
$0.6 $ & $0.997438-0.457833 i$ & $1.504869-0.442113 i$ & $1.121902-0.372080 i$ & $1.500388-0.424225 i$ \\
$0.66$ & $0.957290-0.439127 i$ & $1.444789-0.423710 i$ & $1.076819-0.355724 i$ & $1.440646-0.406203 i$ \\
\hline
\end{tabular*}
\caption{Dominant ($n=0$) quasinormal modes of the scalar and electromagnetic fields for the Hayward black hole ($D=7$, $r_0=1$) calculated using the sixth order WKB formulas with the Padé approximant ($\widetilde{m}=4$).}
\end{table*}

\begin{table*}
\begin{tabular*}{\textwidth}{@{\extracolsep{\fill}} l c c c c}
\hline
$\alpha$ & scalar field ($\ell=0$) & scalar field ($\ell=1$) & $V_1$ ($\ell=1$) & $V_2$ ($\ell=1$) \\
\hline
$0   $ & $1.668942-0.784260 i$ & $2.320534-0.760343 i$ & $1.883886-0.643983 i$ & $2.292528-0.751149 i$ \\
$0.1 $ & $1.638830-0.743134 i$ & $2.277900-0.723564 i$ & $1.913170-0.634942 i$ & $2.256823-0.714800 i$ \\
$0.2 $ & $1.600256-0.701983 i$ & $2.227565-0.686962 i$ & $1.487985-0.363511 i$ & $2.213444-0.677413 i$ \\
$0.3 $ & $1.552660-0.663907 i$ & $2.168898-0.651829 i$ & $1.621770-0.517020 i$ & $2.161046-0.640270 i$ \\
$0.4 $ & $1.498059-0.631064 i$ & $2.101577-0.618607 i$ & $1.599535-0.501117 i$ & $2.098547-0.604240 i$ \\
$0.5 $ & $1.438112-0.601114 i$ & $2.024465-0.586785 i$ & $1.548089-0.472289 i$ & $2.024452-0.569773 i$ \\
$0.6 $ & $1.371496-0.570760 i$ & $1.934580-0.555137 i$ & $1.478035-0.441605 i$ & $1.935998-0.536580 i$ \\
$0.7 $ & $1.293392-0.537224 i$ & $1.825696-0.521669 i$ & $1.392431-0.412445 i$ & $1.827461-0.503190 i$ \\
$0.71$ & $1.284631-0.533567 i$ & $1.813347-0.518106 i$ & $1.382963-0.409587 i$ & $1.815106-0.499736 i$ \\
\hline
\end{tabular*}
\caption{Dominant ($n=0$) quasinormal modes of the scalar and electromagnetic fields for the Hayward black hole ($D=8$, $r_0=1$) calculated using the sixth order WKB formulas with the Padé approximant ($\widetilde{m}=4$).}\label{table:Haywardf}
\end{table*}

\begin{figure*}
\centerline{\resizebox{\linewidth}{!}{\includegraphics{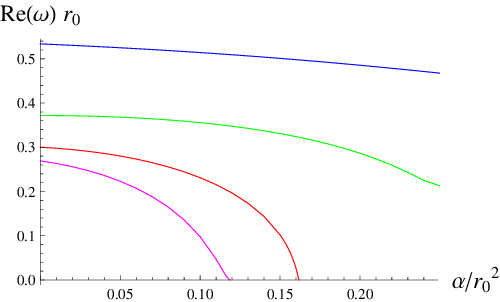}\includegraphics{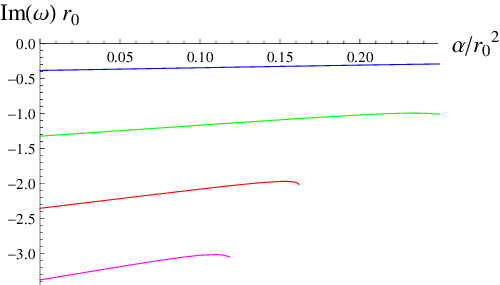}}}
\caption{Dominant quasinormal modes of the scalar field ($\ell=0$) for the Hayward black hole ($D=5$): $n=0$ (blue, top), $n=1$ (green), $n=2$ (red), $n=3$ (magenta, bottom).}\label{fig:D5s0l0}
\end{figure*}

\begin{figure*}
\centerline{\resizebox{\linewidth}{!}{\includegraphics{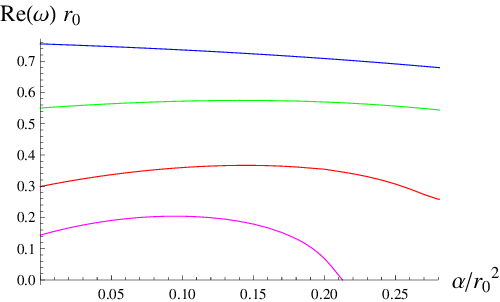}\includegraphics{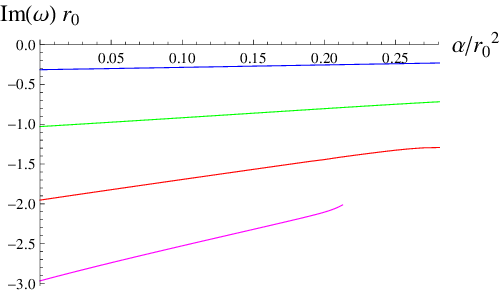}}}
\centerline{\resizebox{\linewidth}{!}{\includegraphics{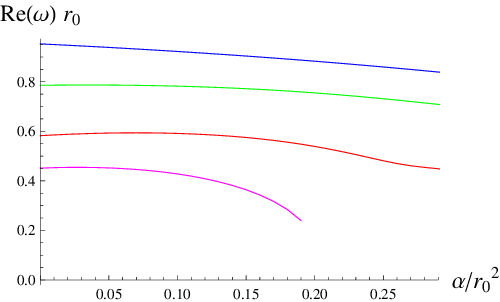}\includegraphics{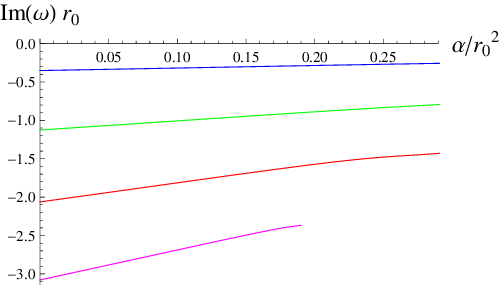}}}
\caption{Dominant quasinormal modes of the electromagnetic field ($\ell=1$), $V_1$ (top panels) and $V_2$ (bottom panels), for the D=5 Hayward black hole: $n=0$ (blue, top), $n=1$ (green), $n=2$ (red), $n=3$ (magenta, bottom).}\label{fig:D5s1l1}
\end{figure*}

\begin{table}
\begin{tabular*}{\linewidth}{@{\extracolsep{\fill}}l c c }
\hline
$\alpha$ & WKB6 Padé & Frobenius \\
\hline
$0  $  & $0.533313-0.383482 i$ & $0.533836-0.383375 i$ \\
$0.1$  & $0.513522-0.344841 i$ & $0.514228-0.344623 i$ \\
$0.2$  & $0.485086-0.307249 i$ & $0.485554-0.306933 i$ \\
$0.3$  & $0.447699-0.276072 i$ & $0.448378-0.275427 i$ \\
$0.4$  & $0.410762-0.252623 i$ & $0.410725-0.252508 i$ \\
\hline
\end{tabular*}
\caption{Quasinormal modes of the scalar field for the Hayward black hole ($D=5$) calculated using the sixth order WKB with Padé approximant $\widetilde{m}=4$ and the accurate Frobenius method: $r_0=1$, $\ell=0$, $n=0$.}\label{table:Haywards0l0check}
\end{table}

From tables~\ref{table:Haywardb}--\ref{table:Haywardf} and Figs.~\ref{fig:D5s0l0}~and~\ref{fig:D5s1l1} we can see that both the real oscillation frequency and the damping rate decrease as the coupling $\alpha$ is increased up to some near critical value for which small nonmonotonic behavior is possible for the damping rate, as can be noticed in Figs.~\ref{fig:D5s0l0}~and~\ref{fig:D5s1l1}. From Table~\ref{table:Haywards0l0check}, we see that the sixth order WKB method with Padé approximants produces sufficiently accurate results even for the worst $\ell=0$ case, once we are limited by the fundamental mode. However, for the overtones, we have to resort to the convergent Frobenius method.
From Figs.~\ref{fig:D5s0l0}~and~\ref{fig:D5s1l1}, we see that while the fundamental mode changes by only a few percents, the real oscillation frequency of the first and a few higher overtones change by a few times, leading even to a qualitatively new behavior when $\re{\omega}$ tends to zero and the mode apparently disappears from the spectrum. Such an outburst of overtones is due to the deformation of the black hole metric near the event horizon when the coupling $\alpha$ is turned on.
It is also worth mentioning that a similar behavior of the vanishing real part of the frequency has recently been observed for the massive scalar field in the background of brane-localized black holes, as discussed in \cite{Zinhailo:2024jzt}.

\begin{table*}
\begin{tabular*}{\linewidth}{@{\extracolsep{\fill}} l | c c | c c }
\hline
$\alpha$\mbox{\hspace{20pt}} & WKB6 Padé ($V_1$) & \mbox{\hspace{30pt}}Frobenius ($V_1$)\mbox{\hspace{30pt}} & WKB6 Padé ($V_2$) & Frobenius ($V_2$) \\
\hline
$0  $ & $0.753155-0.317570 i$ & $0.755414-0.315465 i$ & $0.952729-0.350732 i$ & $0.952728-0.350739 i$ \\
$0.1$ & $0.736623-0.286883 i$ & $0.736044-0.286128 i$ & $0.922189-0.317769 i$ & $0.922129-0.317794 i$ \\
$0.2$ & $0.709222-0.256504 i$ & $0.708579-0.255546 i$ & $0.882942-0.284722 i$ & $0.882831-0.284745 i$ \\
$0.3$ & $0.672309-0.227728 i$ & $0.671503-0.226158 i$ & $0.833939-0.253698 i$ & $0.833832-0.253814 i$ \\
$0.4$ & $0.625649-0.202608 i$ & $0.625005-0.201118 i$ & $0.775548-0.226912 i$ & $0.775337-0.227031 i$ \\
\hline
\end{tabular*}
\caption{Quasinormal modes of the electromagnetic field for the Hayward black hole calculated using the sixth order WKB with Padé approximant $\widetilde{m}=4$ and the accurate Frobenius method; $r_0=1$, $\ell=1$, $n=0$, $D=5$.}\label{table:Haywards1l1check}
\end{table*}

At the same time, an important technical observation we have made is that the conventional WKB method, even up to the sixth order -– typically considered the most accurate -– proves insufficiently precise, even for $\ell=1$, $n=0$, unless supplemented by Padé approximants. For instance, the disparity between the conventional WKB data and the WKB data incorporating Padé approximants can amount to tens of percent, whereas the latter only deviates from the exact Frobenius results by a fraction of one percent (see Table~\ref{table:Haywards1l1check}). Therefore, we conclude that the sixth order WKB formula with the Padé approximant is suitable for obtaining sufficiently accurate values of the quasinormal modes of the Hayward black holes. In the ancillary \emph{Mathematica\textregistered} file\footnote{The \emph{Mathematica\textregistered} file is available from \url{https://arxiv.org/src/2403.07848v2/anc}.}, we share the values of the dominant modes for $5\leq D\leq 11$.

We can see the qualitatively similar behavior for higher multipole numbers from the expressions for quasinormal frequencies in the eikonal ($\ell\to\infty$) limit. Following \cite{Konoplya:2023moy}, we define
$$\kappa=\ell+\frac{1}{2}(D-3),$$
and expand the effective potential in terms of $\alpha$. Then, from the first-order WKB formula, we obtain (cf.~\cite{Konoplya:2003ii})
\begin{eqnarray}\nonumber
\omega&=&\frac{\kappa}{R_s} \left(1+\frac{2 \alpha }{(D-3)^2R_s^2}+\frac{2(2 D+1) \alpha ^2}{(D-3)^4 R_s^4}\right)
\\&&\label{Haywardeikonal}
-i\left(n+\frac{1}{2}\right)\frac{\sqrt{D-3}}{R_s}\Biggl(1-\frac{2(D-2)\alpha}{(D-3)^2R_s^2}
\\\nonumber&&
-\frac{2 \left(3 D^2+2 D+4\right) \alpha ^2}{(D-3)^4 R_s^4}\Biggr) + \Order{\kappa^{-1},\alpha^3},
\end{eqnarray}
where we have introduced
\begin{equation}
R_s=\sqrt{\frac{D-1}{D-3}}\left(\frac{D-1}{2}\mu\right)^{\frac{1}{D-3}}.
\end{equation}

The formula (\ref{Haywardeikonal}) provides quite a good approximation for the dominant modes of the scalar and electromagnetic ($V_2$) fields already for $\ell\geq2$. For the other electromagnetic polarization ($V_1$) the eikonal formula is accurate only for sufficiently large $\ell$. We believe that the approximation can be improved by taking into account the terms beyond the eikonal limit. However, the analysis of such terms is beyond the scope of the present work.

\section{Quasinormal modes of the D-dimensional Bardeen-like black holes}\label{sec:Bardeen}

The $D$-dimensional Bardeen-like black hole is obtained by choosing \cite{Bueno:2024dgm}
\begin{equation}
h(\psi)=\dfrac{\psi}{\sqrt{1-\alpha^2\psi^2}},
\end{equation}
so that the coefficients in (\ref{hseries}) are chosen as
$$\alpha_n=\frac{(1-(-1)^n)\Gamma \left(\frac{n}{2}\right)}{2\sqrt{\pi } \Gamma \left(\frac{n+1}{2}\right)}\alpha^{n-1}.$$

Then the metric function takes the following form:
\begin{equation}\label{Bardeenmetric}
f(r)=1-\dfrac{\mu r^2}{\sqrt{r^{2 (D-1)}+\alpha ^2 \mu^2}}.
\end{equation}

Again, we fix the constant $\mu$ in units of the event horizon as
\begin{equation}
\mu=\dfrac{r_0^{D-1}}{\sqrt{r_0^4 - \alpha^2}}, \qquad 0\leq\alpha\leq\sqrt{\dfrac{D-3}{D-1}}r_0^2.
\end{equation}

\begin{figure*}
\centerline{\resizebox{\linewidth}{!}{\includegraphics{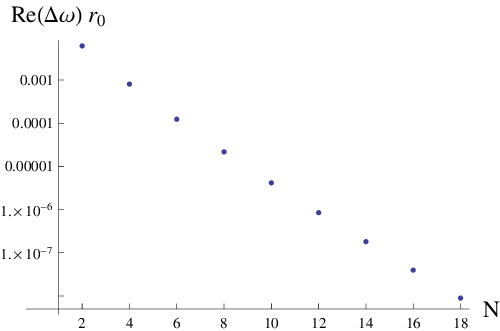}\includegraphics{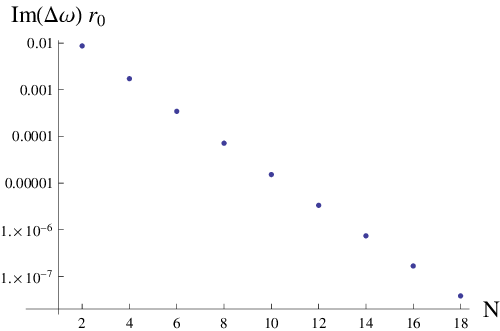}}}
\caption{Convergence of the approximated Frobenius modes of the $\ell=n=0$ scalar field for the Bardeen-like black hole ($D=6$, $\alpha=0.5r_0^2$): Difference between the approximation by $N$ and $N+2$ terms approaches zero for large $N$.}\label{fig:Bardeenerror}
\end{figure*}

\begin{table}
\begin{tabular*}{\linewidth}{@{\extracolsep{\fill}} l c c}
\hline
$\alpha$ & WKB6 Padé & Frobenius \\
\hline
$0  $ & $0.889368-0.532938 i$ & $0.889440-0.533099 i$ \\
$0.1$ & $0.887793-0.530506 i$ & $0.887858-0.530595 i$ \\
$0.2$ & $0.882812-0.523575 i$ & $0.882801-0.523083 i$ \\
$0.3$ & $0.877348-0.505622 i$ & $0.873279-0.510663 i$ \\
$0.4$ & $0.856523-0.491524 i$ & $0.857543-0.493955 i$ \\
$0.5$ & $0.832193-0.474787 i$ & $0.833494-0.475104 i$ \\
$0.6$ & $0.802351-0.459097 i$ & $0.801825-0.458251 i$ \\
$0.7$ & $0.767434-0.441219 i$ & $0.766962-0.440699 i$ \\
\hline
\end{tabular*}
\caption{Quasinormal modes of the scalar field for the Bardeen-like black hole ($D=6$) calculated using the sixth order WKB with Padé approximant $\widetilde{m}=4$ and the accurate Frobenius method: $r_0=1$, $\ell=0$, $n=0$.}\label{table:Bardeens0l0check}
\end{table}

Since the metric function for the Bardeen-like black hole is not a rational function, in order to employ the Frobenius method we approximate the metric function by expanding $f(r)$ in terms of the small coupling parameter $\alpha$ and taking a finite number of terms $N$. This approach is equivalent to a consideration of the finite number of terms in (\ref{hseries}), $n_{max}=N$. In order to check convergence, we have calculated the same mode using different values of $N$ and check that the difference $\Delta\omega\equiv\omega_N-\omega_{N+2}$ quickly approaches zero. The convergence is fast (see Fig.~\ref{fig:Bardeenerror}) and the modes quickly approach the accurate value, for which the sixth order WKB formula with Padé approximant correctly reproduces three decimal places (see Table.~\ref{table:Bardeens0l0check}). We therefore conclude that the numerical values obtained with the Padé approximant provide good practical estimations for the quasinormal modes. Notice that the WKB method is applied to the regular black hole for which infinite number of terms in (\ref{hseries}) are taken into account.

The above observation has two important implications. We see that the dominant quasinormal modes for the black-hole solution, obtained by taking into account the lower-order curvature corrections only, are the same as the corresponding modes of the regular black holes for an infinite tower of the curvature corrections \cite{Bueno:2024dgm}. Therefore, we conclude that observations of the quasinormal ringing cannot probe the regularity of the black-hole solution. The second essential conclusion is that the first few orders of the expansion in powers of the coupling constant provide a reasonable approximation to the basic observational characteristics of black holes, such as quasinormal modes. This observation justifies numerous considerations of quasinormal modes in the theories with particular finite order expansions, such as Lovelock and Gauss-Bonnet theories (see, for example, \cite{Konoplya:2020bxa,Yoshida:2015vua,Zinhailo:2019rwd,Paul:2023eep} and references therein).

The observation mentioned above regarding the proximity of quasinormal modes for black holes, obtained with corrections up to the first few orders versus those for the complete regular metric obtained via an infinite tower of corrections, evidently must also apply to other observable quantities. These may include lensing parameters, quasi-periodic oscillations, and shadows, the latter due to the correspondence between null geodesics and eikonal quasinormal modes \cite{Cardoso:2008bp,Konoplya:2022gjp,Konoplya:2017wot}.

\begin{figure}
\resizebox{\linewidth}{!}{\includegraphics{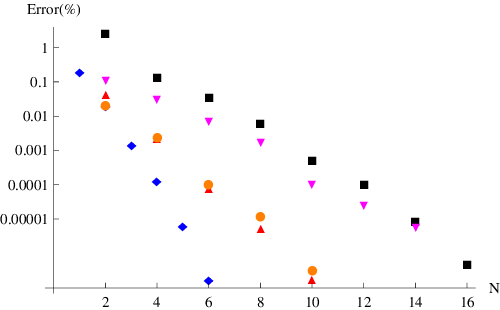}}
\caption{Relative error (in percent) of the value of the dominant quasinormal mode of the scalar field ($\ell=2$) for various regular black holes ($D=5$), approximated by $N$ terms:}
\begin{itemize}
\item[${}_\blacksquare$] $f(r)=1-\dfrac{\mu r^2}{r^{D-1}+\alpha\mu}$ (Hayward) \\ for $\alpha=0.25\mu^2$ (black squares),
\item[${}_\blacklozenge$] $f(r)=1-\dfrac{r^2}{\alpha}\left(1-e^{\displaystyle-\alpha\mu/r^{D-1}}\right)$ \\ for $\alpha=0.25\mu^2$ (blue diamonds),
\item[$\blacktriangle$] $f(r)=1-\dfrac{2\mu r^2}{r^{D-1}+2\alpha\mu+\sqrt{r^{2(D-1)}+4\alpha\mu r^{D-1}}}$ \\ for $\alpha=0.05\mu^2$ (red up-pointing triangles),
\item[$\blacktriangledown$] $f(r)=1-\dfrac{2\mu r^2}{r^{D-1}+\sqrt{r^{2(D-1)}+4\alpha^2\mu^2}}$ for $\alpha=0.25\mu^2$\\(magenta down-pointing triangles),
\item $f(r)=1-\dfrac{\mu r^2}{\sqrt{r^{2(D-1)}+\alpha^2\mu^2}}$ (Bardeen-like) \\ for $\alpha=0.25\mu^2$ (orange circles).
\end{itemize}
\label{fig:convergence}
\end{figure}

Hence, this observation stands as the most significant result of our work. The natural question arises: Does it only apply to the Bardeen metric, or is it a general phenomenon? To ascertain the universality of this observation, we compare the values calculated by the sixth order WKB method with Padé approximants for all five regular black hole metrics derived in \cite{Bueno:2024dgm} for $\ell=2$ scalar perturbations, as an illustration. Fig.~\ref{fig:convergence} demonstrates that, for the considered parameters, even the first corrections yield a relative error within a small fraction of one percent. The relative error becomes larger as the expansion parameter (coupling constant) increases. Nevertheless, we observe the quick convergence of the quasinormal mode with respect to $N$.

\begin{table*}
\begin{tabular*}{\textwidth}{@{\extracolsep{\fill}} l c c c c}
\hline
$\alpha$ & scalar field ($\ell=0$) & scalar field ($\ell=1$) & $V_1$ ($\ell=1$) & $V_2$ ($\ell=1$) \\
\hline
$0  $ & $0.533313-0.383482 i$ & $1.016023-0.362324 i$ & $0.753155-0.317570$ i & $0.952729-0.350732 i$ \\
$0.1$ & $0.531847-0.380844 i$ & $1.013460-0.360304 i$ & $0.752125-0.316798$ i & $0.950776-0.348650 i$ \\
$0.2$ & $0.528364-0.374375 i$ & $1.005287-0.354323 i$ & $0.748352-0.311494$ i & $0.944429-0.342432 i$ \\
$0.3$ & $0.521732-0.355720 i$ & $0.990715-0.344737 i$ & $0.739855-0.302329$ i & $0.932715-0.332219 i$ \\
$0.4$ & $0.504574-0.341758 i$ & $0.968787-0.332376 i$ & $0.726140-0.289623$ i & $0.914260-0.318597 i$ \\
$0.5$ & $0.483434-0.327118 i$ & $0.938798-0.318268 i$ & $0.705780-0.273850$ i & $0.887628-0.302794 i$ \\
$0.6$ & $0.459207-0.313154 i$ & $0.899793-0.302778 i$ & $0.677313-0.256347$ i & $0.851507-0.286100 i$ \\
$0.7$ & $0.431968-0.295691 i$ & $0.849026-0.284961 i$ & $0.639158-0.238986$ i & $0.803595-0.268483 i$ \\
\hline
\end{tabular*}
\caption{Dominant ($n=0$) Quasinormal modes of the scalar and electromagnetic fields for the Bardeen-like black hole ($D=5$, $r_0=1$) calculated using the sixth order WKB formulas with the Padé approximant ($\widetilde{m}=4$).}\label{table:Bardeenb}
\end{table*}

\begin{table*}
\begin{tabular*}{\textwidth}{@{\extracolsep{\fill}} l c c c c}
\hline
$\alpha$ & scalar field ($\ell=0$) & scalar field ($\ell=1$) & $V_1$ ($\ell=1$) & $V_2$ ($\ell=1$) \\
\hline
$0   $ & $0.889368-0.532938 i$ & $1.446598-0.509220 i$ & $1.047467-0.434546 i$ & $1.400026-0.498177 i$ \\
$0.1 $ & $0.887793-0.530506 i$ & $1.444057-0.507282 i$ & $1.044769-0.432564 i$ & $1.397966-0.496106 i$ \\
$0.2 $ & $0.882812-0.523575 i$ & $1.435691-0.501144 i$ & $1.038236-0.428655 i$ & $1.391189-0.489756 i$ \\
$0.3 $ & $0.877348-0.505622 i$ & $1.420879-0.491274 i$ & $1.030445-0.422186 i$ & $1.378874-0.479239 i$ \\
$0.4 $ & $0.856523-0.491524 i$ & $1.398580-0.478388 i$ & $1.019034-0.411643 i$ & $1.359700-0.465006 i$ \\
$0.5 $ & $0.832193-0.474787 i$ & $1.367978-0.463507 i$ & $1.001857-0.397112 i$ & $1.332151-0.448028 i$ \\
$0.6 $ & $0.802351-0.459097 i$ & $1.327976-0.446989 i$ & $0.976852-0.379010 i$ & $1.294524-0.429333 i$ \\
$0.7 $ & $0.767434-0.441219 i$ & $1.275562-0.427896 i$ & $0.940844-0.358343 i$ & $1.243898-0.409066 i$ \\
$0.77$ & $0.737964-0.424787 i$ & $1.227682-0.411548 i$ & $0.906119-0.343144 i$ & $1.197250-0.392951 i$ \\
\hline
\end{tabular*}
\caption{Dominant ($n=0$) Quasinormal modes of the scalar and electromagnetic fields for the Bardeen-like black hole ($D=6$, $r_0=1$) calculated using the sixth order WKB formulas with the Padé approximant ($\widetilde{m}=4$).}
\end{table*}

\begin{table*}
\begin{tabular*}{\textwidth}{@{\extracolsep{\fill}} l c c c c}
\hline
$\alpha$ & scalar field ($\ell=0$) & scalar field ($\ell=1$) & $V_1$ ($\ell=1$) & $V_2$ ($\ell=1$) \\
\hline
$0   $ & $1.270695-0.665502 i$ & $1.881482-0.641045 i$ & $1.396098-0.525823 i$ & $1.845853-0.630962 i$ \\ $0.1 $ & $1.268829-0.663047 i$ & $1.878718-0.639001 i$ & $1.391245-0.526462 i$ & $1.843552-0.628812 i$ \\ $0.2 $ & $1.262903-0.655812 i$ & $1.870165-0.632883 i$ & $1.382012-0.531282 i$ & $1.836359-0.622426 i$ \\ $0.3 $ & $1.258851-0.646843 i$ & $1.854999-0.622871 i$ & $1.374964-0.535106 i$ & $1.823434-0.611762 i$ \\ $0.4 $ & $1.234485-0.624964 i$ & $1.832143-0.609754 i$ & $1.365993-0.532606 i$ & $1.803457-0.597294 i$ \\ $0.5 $ & $1.208145-0.607569 i$ & $1.800732-0.594621 i$ & $1.351809-0.524739 i$ & $1.774905-0.579928 i$ \\ $0.6 $ & $1.174627-0.590892 i$ & $1.759635-0.577852 i$ & $1.330253-0.512522 i$ & $1.735987-0.560535 i$ \\ $0.7 $ & $1.134138-0.572626 i$ & $1.705665-0.558438 i$ & $1.297435-0.495866 i$ & $1.683461-0.539121 i$ \\ $0.8 $ & $1.082436-0.547658 i$ & $1.630504-0.533276 i$ & $1.244634-0.473285 i$ & $1.609427-0.513586 i$ \\ $0.81$ & $1.076195-0.544523 i$ & $1.621150-0.530207 i$ & $1.237593-0.470552 i$ & $1.600195-0.510604 i$ \\
\hline
\end{tabular*}
\caption{Dominant ($n=0$) Quasinormal modes of the scalar and electromagnetic fields for the Bardeen-like black hole ($D=7$, $r_0=1$) calculated using the sixth order WKB formulas with the Padé approximant ($\widetilde{m}=4$).}
\end{table*}

\begin{table*}
\begin{tabular*}{\textwidth}{@{\extracolsep{\fill}} l c c c c}
\hline
$\alpha$ & scalar field ($\ell=0$) & scalar field ($\ell=1$) & $V_1$ ($\ell=1$) & $V_2$ ($\ell=1$) \\
\hline
$0   $ & $1.668942-0.784260 i$ & $2.320534-0.760343 i$ & $1.883886-0.643983 i$ & $2.292528-0.751149 i$ \\
$0.1 $ & $1.666827-0.781897 i$ & $2.317714-0.758372 i$ & $1.881243-0.615438 i$ & $2.290105-0.749094 i$ \\
$0.2 $ & $1.660201-0.774818 i$ & $2.308858-0.752618 i$ & $1.805462-0.573794 i$ & $2.282373-0.743136 i$ \\
$0.3 $ & $1.648428-0.765112 i$ & $2.293407-0.742639 i$ & $1.753821-0.613950 i$ & $2.268966-0.732498 i$ \\
$0.4 $ & $1.629763-0.745468 i$ & $2.270047-0.729480 i$ & $1.740449-0.631430 i$ & $2.248263-0.718004 i$ \\
$0.5 $ & $1.601906-0.728127 i$ & $2.237870-0.714326 i$ & $1.725736-0.632044 i$ & $2.218763-0.700581 i$ \\
$0.6 $ & $1.565776-0.711031 i$ & $2.195736-0.697632 i$ & $1.703628-0.624599 i$ & $2.178683-0.681056 i$ \\
$0.7 $ & $1.521195-0.692466 i$ & $2.140348-0.678328 i$ & $1.669943-0.611112 i$ & $2.124612-0.659294 i$ \\
$0.8 $ & $1.463718-0.667690 i$ & $2.063055-0.653194 i$ & $1.615919-0.590095 i$ & $2.048149-0.633088 i$ \\
$0.84$ & $1.433891-0.654268 i$ & $2.021402-0.639942 i$ & $1.584189-0.578309 i$ & $2.006807-0.620012 i$ \\
\hline
\end{tabular*}
\caption{Dominant ($n=0$) Quasinormal modes of the scalar and electromagnetic fields for the Bardeen-like black hole ($D=8$, $r_0=1$) calculated using the sixth order WKB formulas with the Padé approximant ($\widetilde{m}=4$).}\label{table:Bardeenf}
\end{table*}

Referring tables~\ref{table:Bardeenb}--\ref{table:Bardeenf}, we observe a decrease in both the real oscillation frequency and damping rate as the coupling is increased. Qualitatively, this behavior of the dominant mode can be explained by the characteristics of the potential peak, which, in the units of the event horizon, becomes lower as $\alpha$ grows. However, in the units of the asymptotic mass, the dependence of the coupling parameter becomes not so straightforward and cannot be attributed to such a simple geometric feature of the effective potential.
Additionally, we note an outburst of overtones in this scenario, although we refrain from presenting numerical data for this case to prevent overloading the manuscript.

\begin{figure*}
\centerline{\resizebox{\linewidth}{!}{\includegraphics{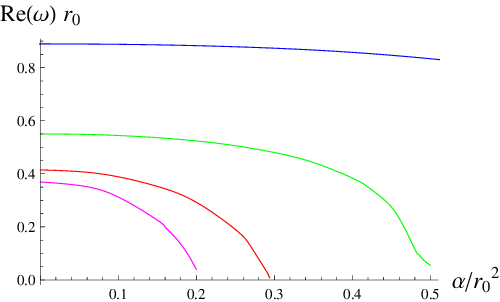}\includegraphics{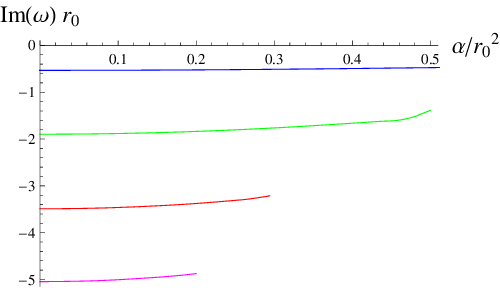}}}
\caption{Dominant quasinormal modes of the scalar field ($\ell=0$) for the Bardeen-like black hole ($D=6$): $n=0$ (blue, top), $n=1$ (green), $n=2$ (red), $n=3$ (magenta, bottom).}\label{fig:BardeenD6s0l0}
\end{figure*}

The qualitative behavior of the overtones of the Bardeen-like black hole is similar to the one of the Hayward black hole: The real part of the quasinormal modes quickly approaches zero when $\alpha$ approaches some critical values (see Fig.~\ref{fig:BardeenD6s0l0}). This is an interesting observation, worthy of further investigation, because if it occurs that similar behavior is appropriate not only for the Bardeen and Hayward metrics, but also for other models with higher-curvature corrections, that may be a general distinctive feature of such models.

Again, using the first-order WKB formula we can derive the eikonal approximation for the quasinormal modes of the Bardeen-like black holes,
\begin{eqnarray}\nonumber
\omega&=&\frac{\kappa}{R_s} \left(1+\frac{2 \alpha^2 }{(D-3)^3R_s^4}+\frac{2(6 D-1) \alpha ^4}{(D-3)^6 R_s^8}\right)
\\&&\label{Bardeeneikonal}
-i\left(n+\frac{1}{2}\right)\frac{\sqrt{D-3}}{R_s}\Biggl(1-\frac{2(3D-4)\alpha^2}{(D-3)^3R_s^4}
\\\nonumber&&
-\frac{2 \left(33 D^2-42 D+4\right) \alpha ^4}{(D-3)^6 R_s^8}\Biggr) + \Order{\kappa^{-1},\alpha^6},
\end{eqnarray}
where
\begin{equation}
\kappa=\ell+\frac{D-3}{2},\quad R_s=\sqrt{\frac{D-1}{D-3}}\left(\frac{D-1}{2}\mu\right)^{\frac{1}{D-3}}.
\end{equation}
The formula (\ref{Bardeeneikonal}) for $\ell\geq2$ produces the best approximation for the quasinormal modes of the scalar field.

The well-known correspondence between the eikonal quasinormal modes of a stationary, spherically symmetric and asymptotically flat black holes and parameters of the circular null geodesic states that the real and imaginary parts of the $\ell \gg n$ quasinormal mode are multiples of the frequency and instability timescale (Lyapunov exponent) of the circular null geodesics, respectively \cite{Cardoso:2008bp}. Although this correspondence has a number of limitations and counterexamples \cite{Konoplya:2022gjp,Konoplya:2017wot,Bolokhov:2023dxq,Konoplya:2020bxa}, here, one can easily check that it takes place for the test fields under consideration.

\section{Late-time tails}\label{sec:tails}

Here, having in mind both metrics (Hayward and Bardeen-like) under consideration and all three types of perturbations (one scalar type and two electromagnetic ones), we will derive the law of decay at asymptotically late times $t\to\infty$.

For both types of black holes, the asymptotic behavior of the tortoise coordinate (\ref{tortoise}) does not depend on $\alpha$ up to the order
\begin{equation}\label{tortoiseasymptotic}
r_*=r-\frac{\mu}{(D-4)r^{D-4}}+\Order{\frac{1}{r^{2D-7}}}.
\end{equation}
After inverting the expansion (\ref{tortoiseasymptotic}),
\begin{equation}
r=r_*+\frac{\mu}{(D-4)r_*^{D-4}}+\Order{\frac{1}{r_*^{2D-7}}},
\end{equation}
and substituting it into Eq.~(\ref{potentials}), we find that the dominant asymptotic behavior of the effective potentials does not depend on the coupling $\alpha$,
\begin{eqnarray}\nonumber
V_p&=&\frac{(\ell+\frac{d}{2}-1)(\ell+\frac{d}{2}-2)}{r_*^2}+\mu\frac{K_p}{r_*^{D-1}}+\Order{\frac{1}{r_*^{2D-4}}},
\\&&p=0,1,2.
\end{eqnarray}

Here $K_p$ is a constant, which also does not depend on $\alpha$. For the scalar field \cite{Cardoso:2003jf},
\begin{subequations}
\begin{equation}
K_0=-\frac{\ell(\ell+D-3)(D-2)}{D-4},
\end{equation}
and for the electromagnetic field, we find,
\begin{eqnarray}
K_1&=&\frac{\ell(\ell+D-3)(D-2)}{D-4}+(D-3)^2,\\
K_2&=&\frac{\ell(\ell+D-3)(D-2)}{D-4}+1.
\end{eqnarray}
\end{subequations}

Therefore, following \cite{Ching:1995tj,Cardoso:2003jf}, we conclude that the late-time tails do not depend on $\alpha$ and coincide with the power-law falloff derived for the Tangherlini black holes \cite{Cardoso:2003jf}:
\begin{eqnarray}
\Psi &\propto& t^{-(2 \ell+ D - 2)}, ~~\quad \text{for odd $D$},\\
\Psi &\propto& t^{-(2 \ell+ 3D - 8)}, \quad \text{for even $D$}.
\end{eqnarray}

\begin{figure*}
\centerline{\resizebox{\linewidth}{!}{\includegraphics{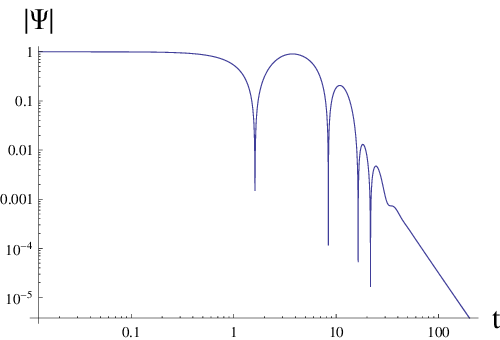}\includegraphics{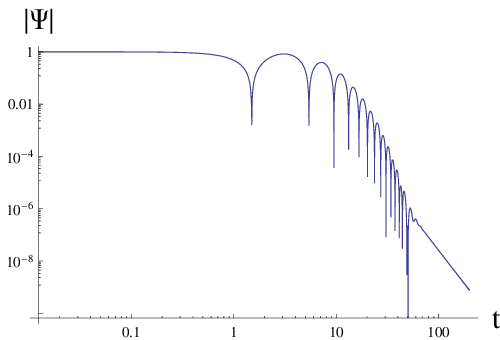}}}
\caption{Late-time decay of a massless scalar field on a logarithmic plot for $\ell=0$ (left) and $\ell=1$ (right) perturbations around the $5$-dimensional Bardeen-like black hole. The decay law is $\propto t^{-3}$ for $\ell=0$ and $\propto t^{-5}$ for $\ell=1$, $\alpha=0.6$. Similar plots with the same decay law can be obtained for the $5$-dimensional Hayward black hole.}\label{fig:TD2}
\end{figure*}

Indeed, in Fig.~\ref{fig:TD2}, we see an example of the above asymptotic decay law for the Bardeen-like black hole at $D=5$. However, for even $D$, the second-order time-domain integration scheme gives an incorrect power-law decay due to the so-called ghost potential (see discussion in the Appendix A of \cite{Ching:1995tj}).

\section{Conclusions}\label{sec:conclusions}

In the present work we calculated quasinormal modes of $D$-dimensional generalization of the Hayward and Bardeen-like black holes, which were obtained as a result of infinite tower of higher-curvature corrections to the Einstein action. We have shown that even at small and moderate values of the coupling constant the overtones deviate at a much stronger rate than the fundamental mode, leading to a qualitatively new features of the spectrum, such as modes with very small (tending to zero) real oscillation frequency. This is demonstration of the phenomenon, known as the outburst of overtones, which happens when the black hole geometry is deformed in the near-horizon zone. Once the near-horizon deformations are relatively small, that is the coupling $\alpha$ is small, the fundamental mode changes only slightly, so that the first few overtones bring the information about the near-horizon geometry.

An important observation we have made during our study is that, when examining theories truncated at different orders of higher-curvature corrections, the dominant quasinormal modes rapidly converge to their limit value with an infinite number of correction terms. This result signifies that considering only the initial correcting orders, as in Gauss-Bonnet and Lovelock gravities, is well-justified as a robust approach to the complete theory.

Our work could be extended by consideration of the massive scalar and vector fields, because, as was recently shown in \cite{Zinhailo:2024jzt,Dubinsky:2024jqi,Konoplya:2024ptj}, the massive term produces qualitatively new behavior at both the quasinormal ringing stage and at asymptotic tails.
However, our primary interest is to generalize the above study of overtones to other
black-hole models with higher-curvature corrections in order to know whether the peculiar overtones' behavior, which takes place for the Bardeen and Hayward metrics, is also appropriate to other higher-curvature corrected black-hole solutions. While in the present work we checked the convergence of the quasinormal frequencies as to the increasing maximal order of higher-curvature corrections for all the five metrics of \cite{Bueno:2024dgm}, no detailed analysis of the quasinormal spectra for the remaining three regular black holes was done.

\begin{acknowledgments}
A.~Z. was supported by Conselho Nacional de Desenvolvimento Científico e Tecnológico (CNPq).
\end{acknowledgments}

\bibliographystyle{unsrt}
\bibliography{bibliography}
\end{document}